\documentclass{moriond}

\def\be{\begin{equation}}
\def\ee{\end{equation}}
\def\bea{\begin{eqnarray}}
\def\eea{\end{eqnarray}}

\newcommand{\lambdabar}{{\mkern0.75mu\mathchar '26\mkern -9.75mu\lambda}}

\newcommand{\Photo}{\includegraphics[height=25mm]{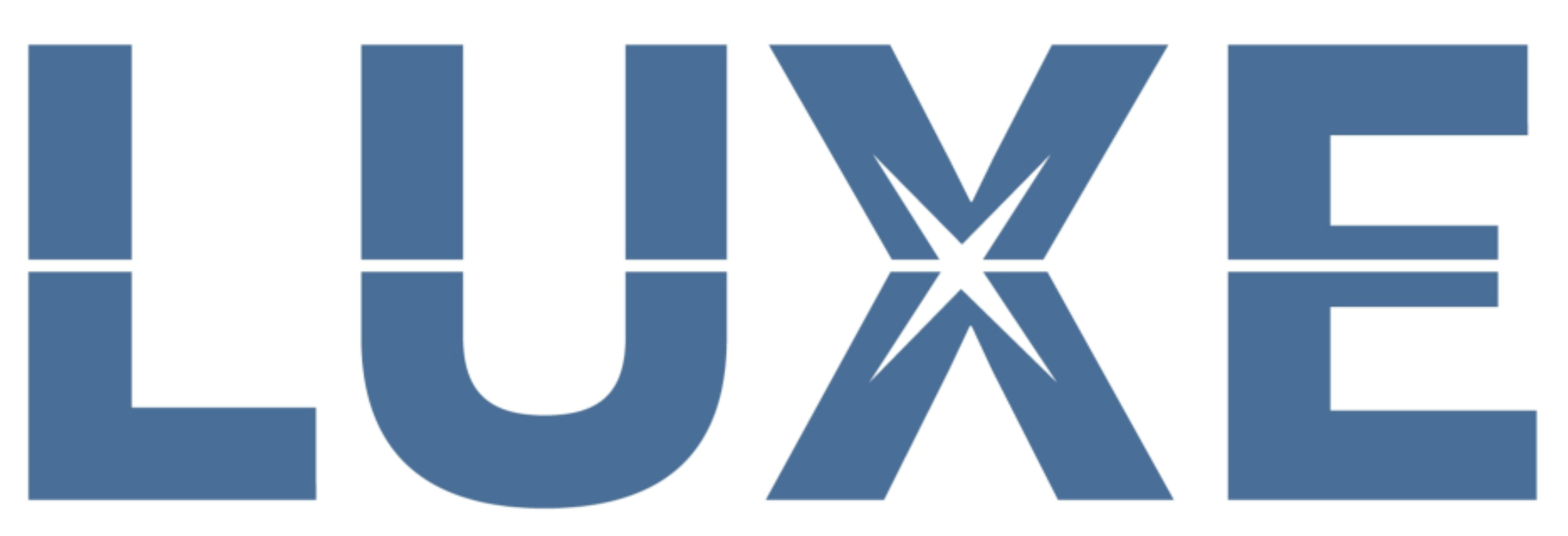}}

\begin{document}
\vspace*{4cm}
\title{Detector Challenges of the strong-field QED experiment LUXE \break at the European XFEL}

\author{ Adri\'an Irles on behalf the LUXE collaboration }

\address{IFIC, Universitat de Val\`encia and CSIC, C./ Catedr\'atico Jos\'e Beltr\'an 2, E-46980 Paterna, Spain}

\maketitle\abstracts{
The LUXE (Laser Und XFEL Experiment) aims at studying high-field QED in electron-laser and photon-laser interactions, with the 16.5 GeV electron beam of the European XFEL and a laser beam with a power of up to 350 TW. The experiment will measure the spectra of electrons, positrons and photons in expected ranges of $10^{-3}$ to $10^9$ per 1 Hz bunch crossing, depending on the laser power and focus. These measurements have to be performed in the presence of a low-energy high radiation background. To meet these challenges, for high-rate electron and photon fluxes, the experiment will make use of dedicated physics-driven detectors at specific locations downstream of the interaction point. 
}

\section{Introduction: QED in strong fields and the LUXE}

For large values of the electromagnetic (EM) field, the Schwinger critical field, 
\begin{equation}
\mathcal{E}_{crit}=\frac{m_{e}^{2}c^{3}}{\hbar e} = 1.32\cdot 10^{18} V/m
\label{eq:murnf}
\end{equation}
is surpassed and the vacuum becomes unstable to electron-positron pair production.

\begin{figure}[h!!]
%\begin{minipage}{0.33\linewidth}
%\centerline{
\centering
\includegraphics[width=0.55\linewidth]{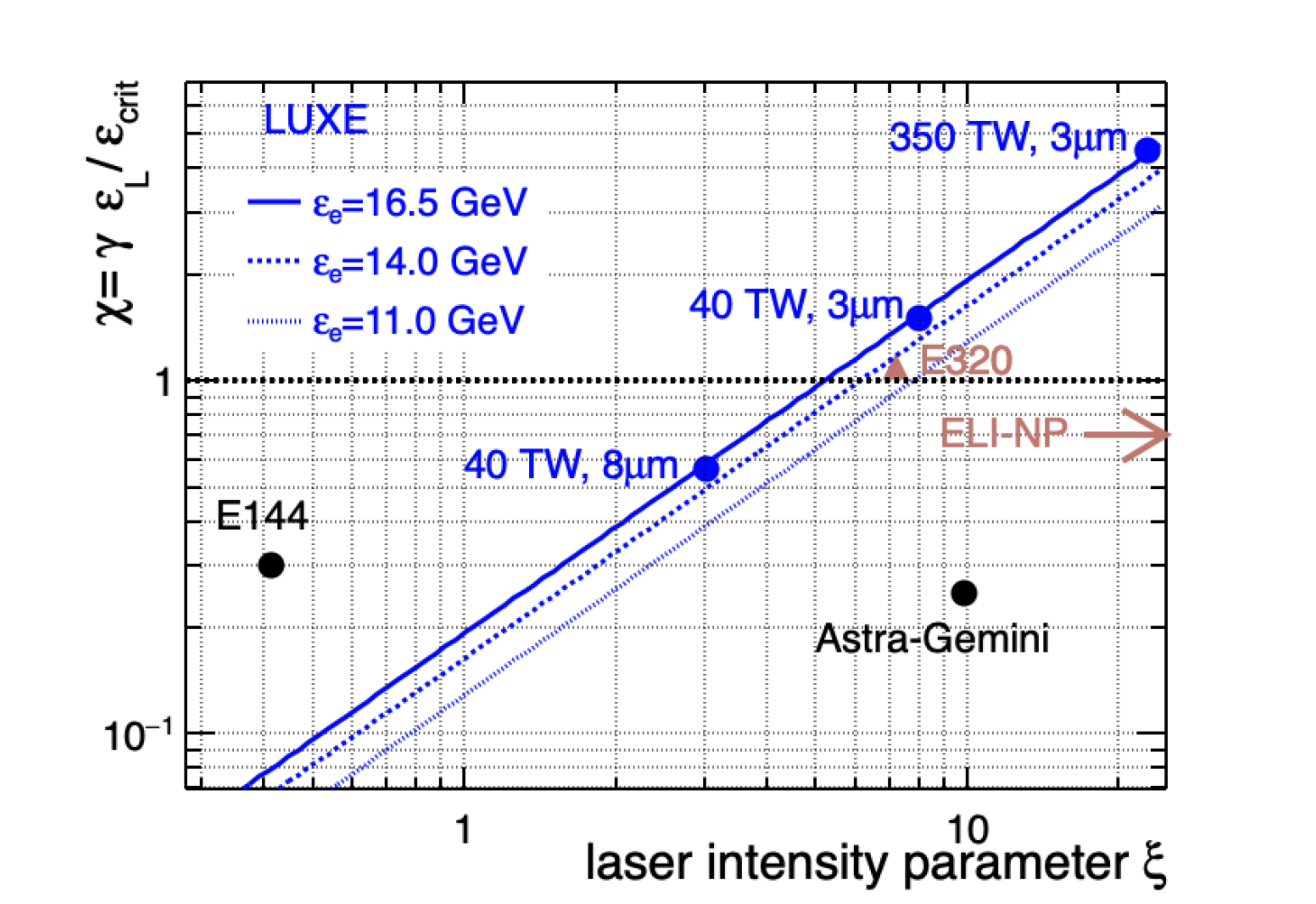} 
%\end{minipage}
\caption[]{Quantum parameter $\chi$ as a function of the intensity parameter $\xi$ for LUXE and a selection of experiments and facilities. Figure extracted from the LUXE Conceptual Design Report \cite{Abramowicz:2021zja}.}
\label{fig:luxe_reach}
\end{figure}

In the presence of such strong fields, Quantum Electrodynamics (QED) becomes non-perturbative \cite{RevModPhys.94.045001,Fedotov:2022ely}. Such strong fields are expected to exist in some extreme environments as on the surface of neutron stars or in the bunches of future linear lepton colliders. Attempts to reach and measure these fields in experiments have been done in the past (E144 \cite{Bamber:1999zt}), and are ongoing in nowadays experiments (AstraGemini, E320 \cite{Clarke:2022rbd,Poder:2017dpw,PhysRevX.8.011020}) or are planned in upcoming high-power laser facilities (ELI-NP \cite{Turcu:2016dxm}).
The LUXE \mbox{experiment \cite{Abramowicz:2021zja}} aims at studying SFQED in electron-laser and photon-laser interactions, with the 16.5 GeV electron beam of the European XFEL and a laser beam with a power of up to 350 TW. A laser with this power cannot reach the Schwinger field strength on its own. Instead, we take advantage of the boost provided when it collides with the high-energy electrons. In the rest of frame system of reference of the electron, the field is enhanced by a factor $\gamma(1+cos\theta)$ with $\gamma$ being the Lorentz factor and $\theta$ the incidence angle between the electron beam and the laser.

In the left plot of Fig. \ref{fig:luxe_reach}, the reach of LUXE at different stages of its operation is compared with the reach of other present and past experiments. 
Two parameters are compared in this figure: the classical non-linearity parameter or laser intensity parameter, $\xi$, and the quantum non-linearity parameter, $\chi$. The former measures the work done by the EM field over an electron Compton wavelength ($\lambdabar=\hbar/(m_{e}c)$) in units of the laser photon energy $\hbar\omega$. Whenever it is larger than unity, the calculation of processes at any given order in the QED coupling, $\alpha$, requires a resummation at all orders of $\xi$. The quantum non-linearity parameter  characterises the field strength experienced by an electron in its rest frame and the recoil experienced by the electron emitting a photon. Both parameters\footnote{The field intensity parameter is defined as $\xi= \frac{m_{e}\mathcal{E}_{L}}{\omega_{L}\mathcal{E}_{crit}}$ where $\omega_{L}$ is the laser wavelength and $\mathcal{E}_{L}$ is the laser electromagnetic field strength. The quantum non-linearity parameter is defined as $\chi=\frac{e\mathcal{E}_{L}\lambdabar}{m_{e}c^{2}}$.} are dependent on the beam and laser parameters.

The main processes of strong-field QED probed by LUXE are non-linear Compton scattering and non-linear Breit-Wheeler pair and nonlinear trident pair creation.
The non-linear Compton scattering occurs when multiple laser photons are absorbed by an electron and a single energetic photon is emitted. This process is 
probed through the measurement of the displacement of the Compton edge as a function of the laser intensity parameter.
In the Breit-Wheeler process, a high-energy photon absorbs multiple laser photons and produces an electron-positron pair. 
The scaling of this process with laser intensity is direct evidence of the transition from perturbative to non-perturbative QED and has no classical equivalent.
More information about the physics process can be found in \cite{Abramowicz:2021zja}.
%, for example, in this proceeding note from Levy A. on behalf LUXE %\cite{Levy:2022qdc}.

\section{Experimental setup}

\begin{figure}[h!]
\centering
\begin{tabular}{cc}
    \includegraphics[width=0.4\textwidth]{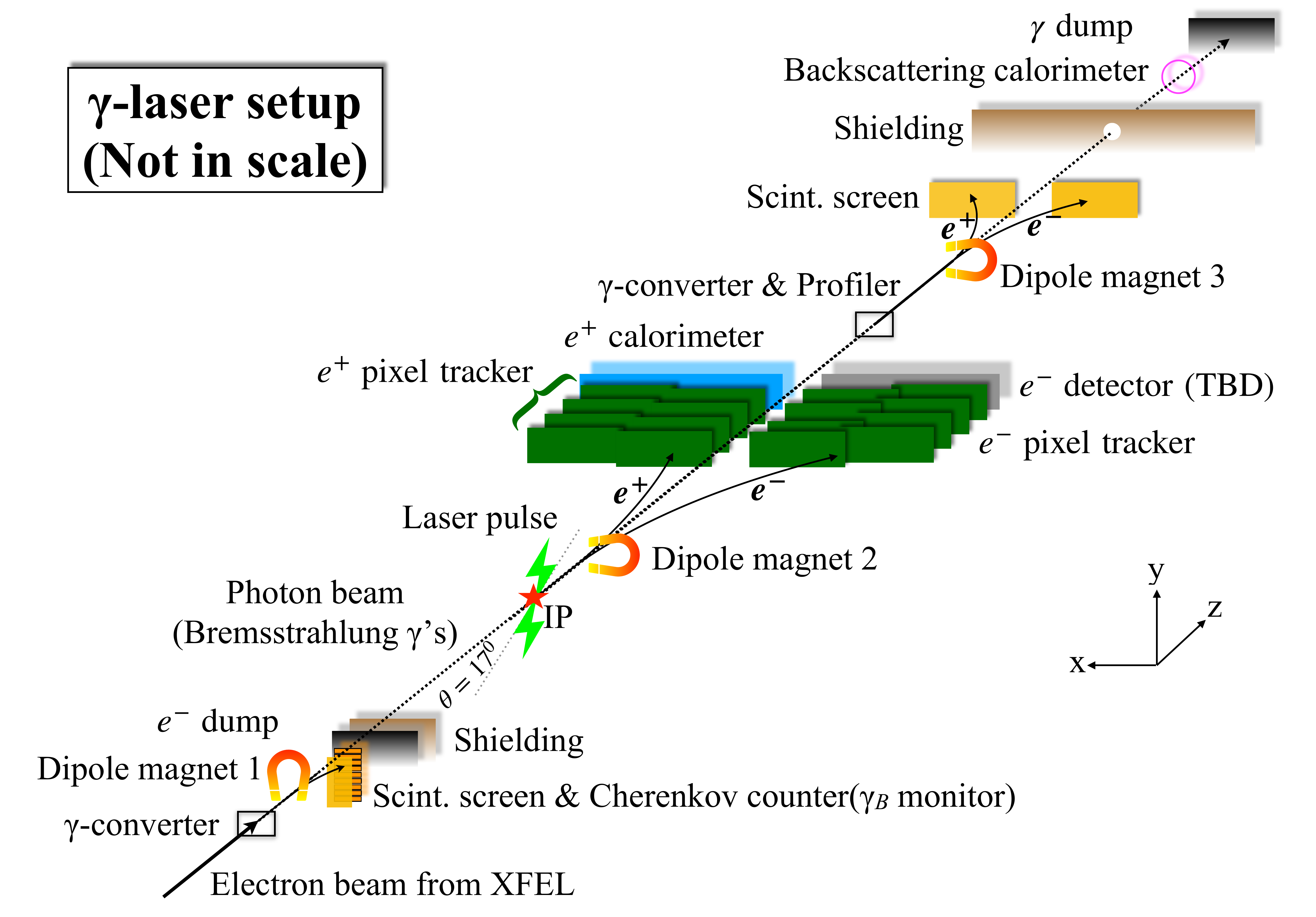} &  \includegraphics[width=0.4\textwidth]{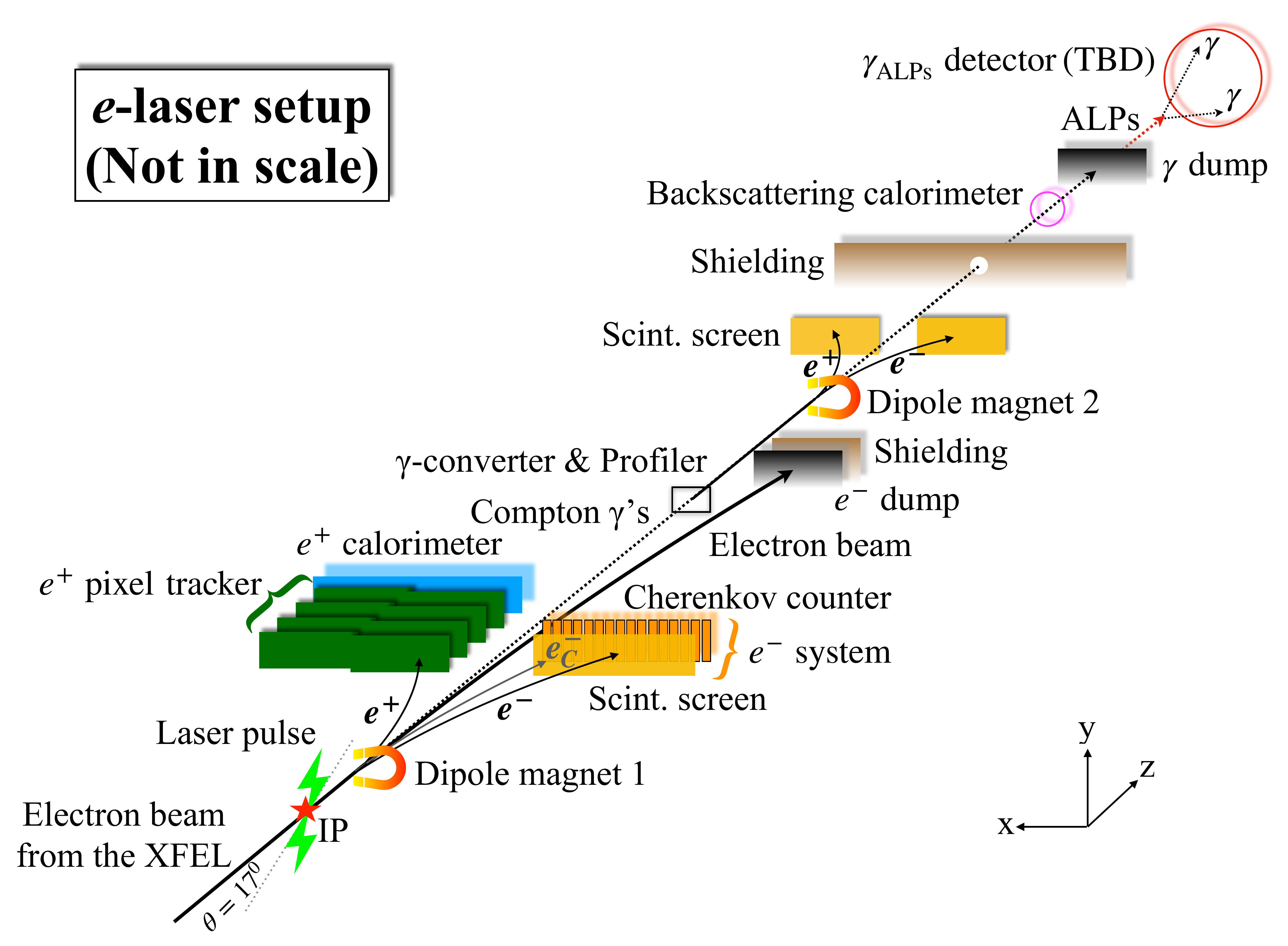} 
\end{tabular}
\caption[]{Sketch of the LUXE experimental setups.}
\label{fig:operation}
\end{figure}

\begin{figure}
%\begin{minipage}{0.33\linewidth}
%\centerline{
\centering
\begin{tabular}{cc}
\includegraphics[width=0.45\linewidth]{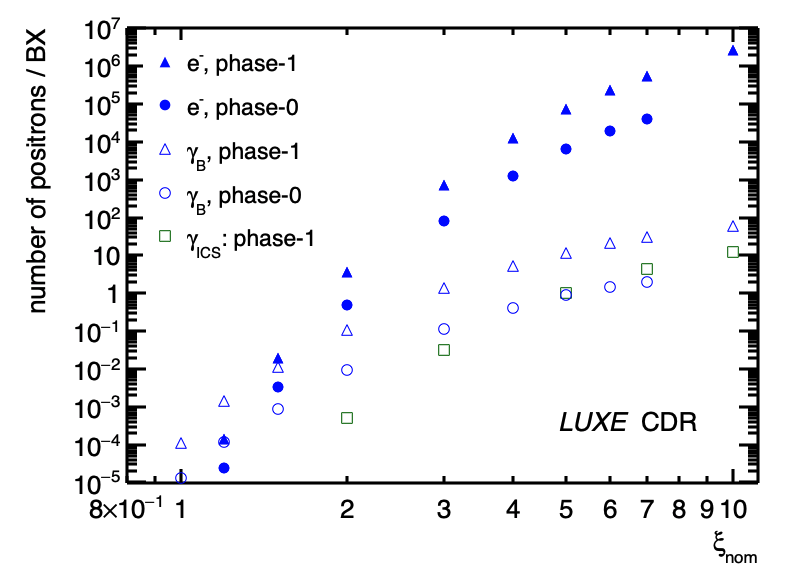}  & \includegraphics[width=0.45\linewidth]{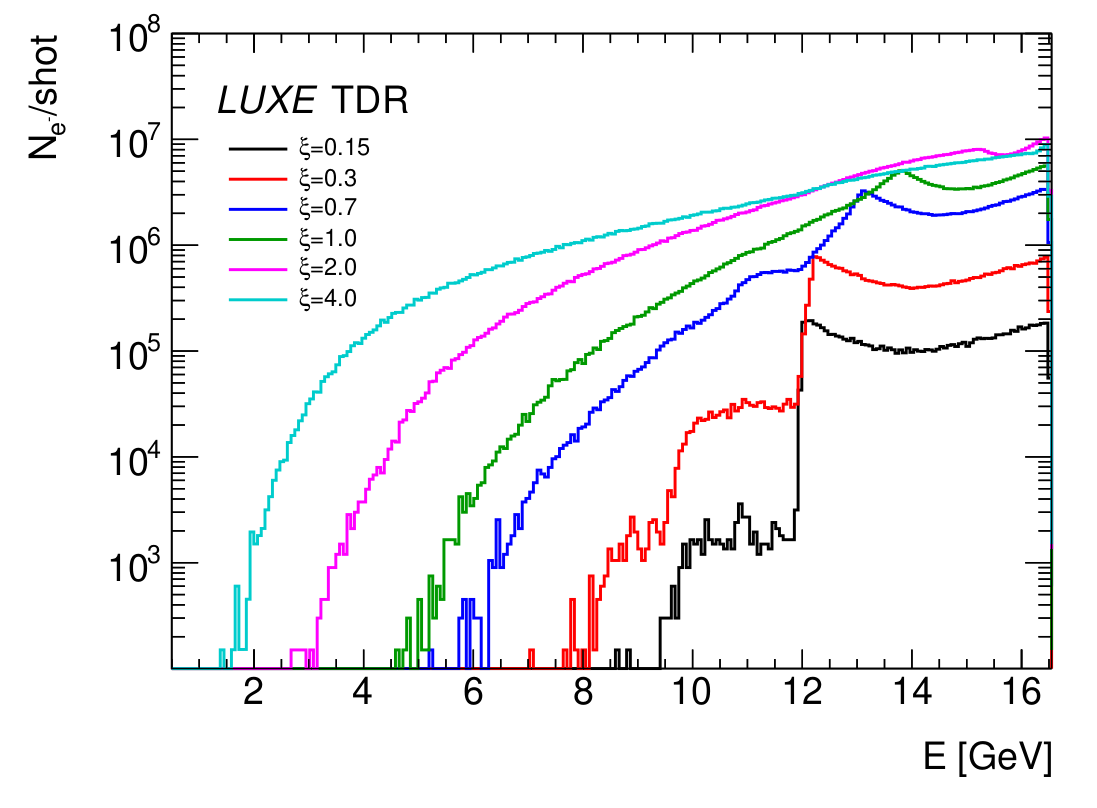}  
\end{tabular}
%\end{minipage}
\caption[]{Left: expected rate of positron production under different running conditions. Right: photon spectra and Compton edge displacement for different beam laser operation points.}
\label{fig:luxe_rates}
\end{figure}

LUXE will run in two modes of operation: the $e$-laser mode in which the intense laser collides directly with the electron beam, and the $\gamma$-laser mode, in which the laser interacts with secondary photons generated by the electron beam in a high-Z target upstream of the interaction point. These two operation modes are shown in Fig. \ref{fig:operation}. The secondary photon beam will be delivered
by impinging the electron beam on a Tungsten Bremsstrahlung target or producing Compton photons after colliding the electron beam with a low-intensity laser pulse.
Depending on the mode of operation, there is a vast range in the multiplicity of the produced electrons, positrons and photons.
This is summarised in Fig. \ref{fig:luxe_rates} for positrons (left) and for photons (right). 
The LUXE collaboration proposes a set of physics-driven detector technologies at specific locations downstream of the interaction point. 
As covering all of them is out of the scope of this contribution we will concentrate on electron 
and positron detection systems for the $e$-laser mode. For a more comprehensive picture, we refer the reader to the Conceptual Design report of LUXE \cite{Abramowicz:2021zja} and the Technical Design report of LUXE which is soon to appear.

\begin{figure}[h!]
\centering
\begin{tabular}{cc}
\includegraphics[width=0.4\textwidth]{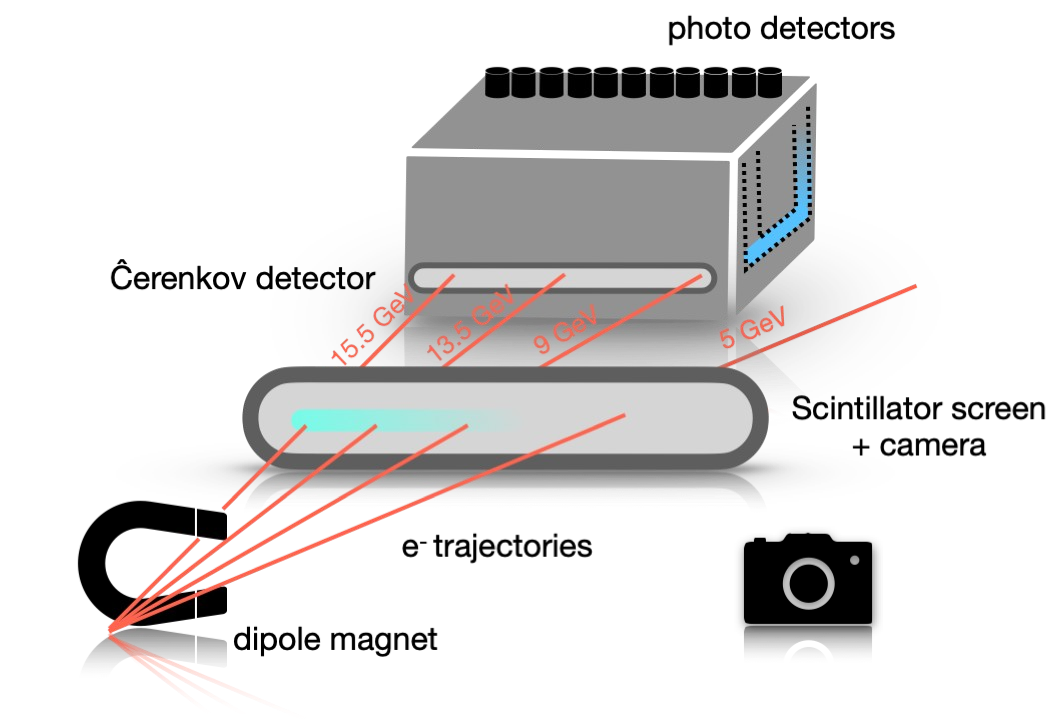} &  \includegraphics[width=0.4\textwidth]{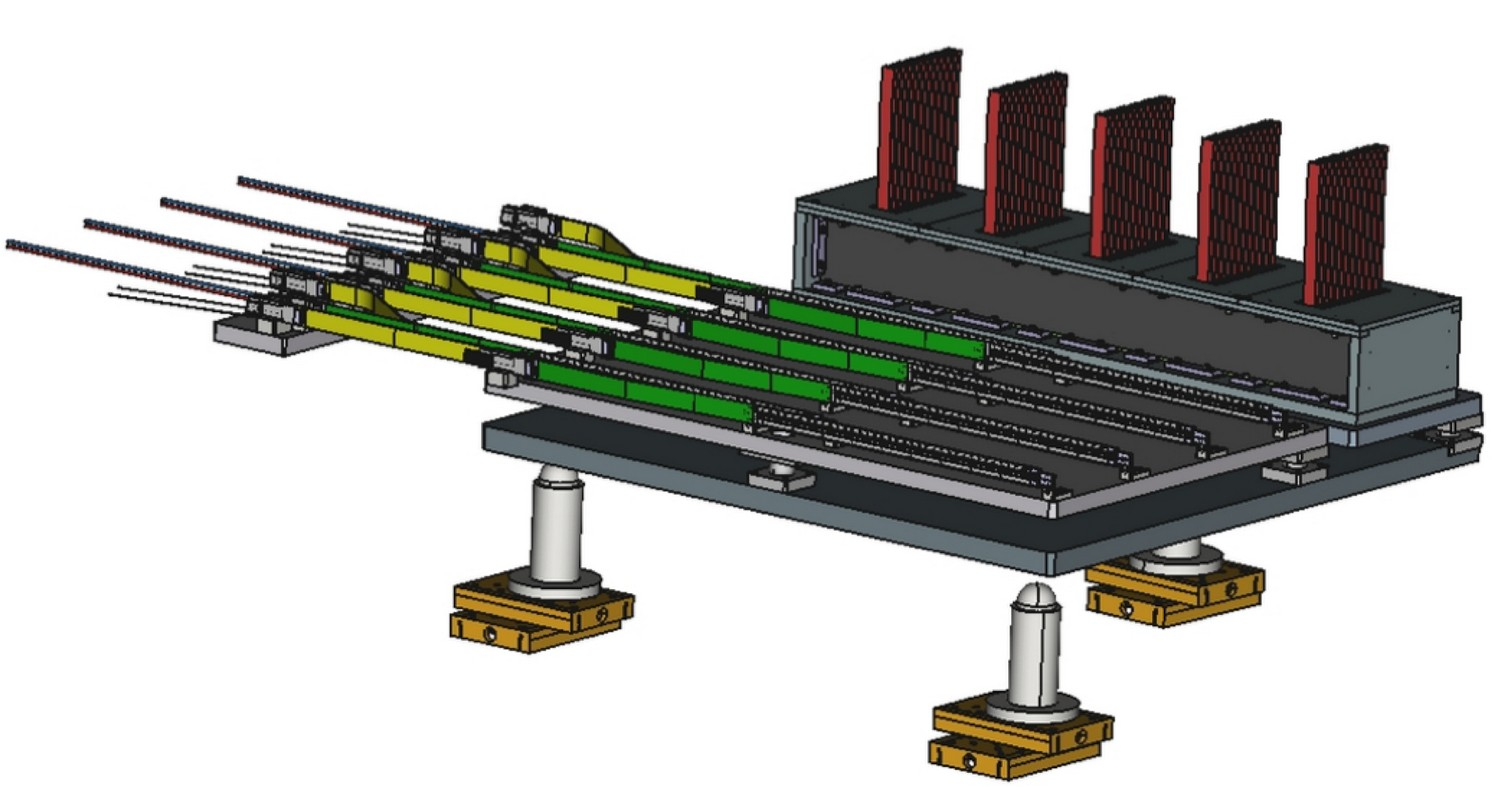}
\end{tabular}
\caption[]{Sketch of the LUXE experimental setups.}
\label{fig:detector}
\end{figure}

%\begin{description}
%\item[
\textbf{Electron Detection System:} The measurement of the non-linear Compton edge requires dealing with very large rates of electrons (up to $10^9$). The solution chosen for this is based on a two-detector system, as shown in the left part of Fig. \ref{fig:detector}: a scintillator screen  and a Cherenkov gas detector. 
The scintillator screen solution has already been used by the AWAKE collaboration at CERN. It features a digital camera that records
the scintillation light with a spatial resolution of 500 $\mu$m. This solution is chosen because of its large signal-over-background ratio (of the order of 100) and its radiation hardness, holding up to 10 MGy.
The Cherenkov detector solution is built on the argon gas polarimeter developed for the International Linear Collider. It features a low refractive index gas that helps to reduce photons and low energy backgrounds due to the low Cherenkov threshold of 20 MeV. It features an excellent signal-over-background ratio larger than 1000. These two detectors fulfil complementary roles, being the Cherenkov detector used to cover the high energy range and the scintillator screen for the low energy.
%\end{description}

\textbf{Positron Detection System:} For the study in detail of the Breit-Wheeler process, the measurement of the positron production rate is crucial. The system will consist of a high-resolution tracker system, with low material budget  and a highly granular electromagnetic calorimeter, as shown in the right part of Fig. \ref{fig:detector}. The tracker will consist of three layers of highly integrated ALPIDE sensors (developed by ALICE for phase 1 upgrade of the LHC) with 5 $\mu$m of spatial resolution and a pixel size of $27\times29$ $\mu$m$^{2}$. The calorimeter system will consist of a multilayer high-granular system developed for the forward calorimetry systems of the future International Linear Collider (ILC). It foresees a depth of 20 layers summing up to 20 radiation lengths provided by 20 tungsten plates of 3.5 mm and a lateral coverage of 810 mm. The sensors (silicon or gallium-arsenide as preferred options) will have a surface of $90\times90$ mm$^{2}$ and will be pixelated in 256 cells of $5.5\times5.5$ mm$^{2}$ each. The setup is required to offer maximal compactness to ensure a minimal Moli\`ere Radius, $R_{M}\approx 3.5$ mm. A dedicated ASIC, FLAXE is being developed for LUXE, by adapting the FLAME ASIC, developed for the ILC. The calorimeter is expected to offer an energy resolution of $\frac{\sigma_{E}}{E}=\frac{19.3\%}{\sqrt{E}}$ and a position resolution of $\sigma_{x}=780$ $\mu$m (calculated for single electrons hitting perpendicular to the first layer).

\section{Conclusions and outlook}

LUXE will likely be the first experiment to take precision measurements in an SFQED regime never explored before in laboratory conditions, where the quantum parameter $\chi > 1$. This will be possible thanks to a high-intensity laser and the European XFEL electron beam. In addition, LUXE requires a set of dedicated detectors able to cope with a vast variety of rates for particles and backgrounds. A sketch of a few of these detector systems has been given in this proceeding. 
LUXE also provides a platform to search for physics beyond the Standard Model, specifically Axion Like Particles, which is discussed in detail here \cite{PhysRevD.106.115034}.

\section*{Acknowledgments}

We thank the DESY technical staff for continuous assistance and the DESY directorate for their strong support and the hospitality they extend to the non-DESY members of the collaboration. This work has benefited from computing services provided by the German National Analysis Facility (NAF) and the Swedish National Infrastructure for Computing (SNIC).
A. Irles acknowledges the financial support from the Generalitat Valenciana (Spain) under grant number CIDEGENT/2020/21 and from the MCIN with funding from the European Union NextGenerationEU and Generalitat Valenciana in the call Programa de Planes Complementarios de I+D+i (PRTR 2022). Project \textit{Si4HiggsFactories}, reference ASFAE$/2022/015$

\section*{References}
%\section*{References}

%\begin{thebibliography}{99}
\bibliographystyle{JHEP}
\bibliography{ref} 

%\end{thebibliography}

\end{document}